\documentclass[showpacs,twocolumn,superscriptaddress,preprintnumbers,fleqn,widetext]{revtex4} 
\usepackage{graphicx}  % needed for figures
\usepackage{dcolumn}   % needed for some tables
\usepackage{bm}        % for math
\usepackage{amssymb}   % for math

\begin{document}

\preprint{DFPD-12/TH/09, ROM2F/2012/09}

\title{Evidence for a family of SO(8) gauged supergravity theories}

\author{G.~Dall'Agata} \affiliation{Dipartimento di Fisica e Astronomia, Universit\`a degli Studi di Padova, Via Marzolo 8, 35131, Padova, Italy}
\affiliation{INFN, Sezione di Padova, Via Marzolo 8, 35131, Padova, Italy}
\author{G.~Inverso} \affiliation{Dipartimento di Fisica, Universit\`a di Roma Tor Vergata} \affiliation{INFN, Sezione di Roma 2, via della Ricerca Scientifica, 00133 Roma, Italy}
\author{M.~Trigiante} \affiliation{Laboratory of Theoretical Physics,
Dipartimento di Scienze Applicate e Tecnologia, Politecnico di Torino, C.so Duca degli Abruzzi, 10129 Torino, Italy} \affiliation{INFN, Sezione di Torino, Via Pietro Giuria 1, 10125, Torino, Italy}

\date{\today}

\begin{abstract}
In this note we discuss the classification of duality orbits of $N=8$ gauged supergravity models.
Using tensor classifiers, we show that there is a one-parameter family of inequivalent SO(8) gauged supergravity theories.
We briefly discuss the couplings of such models and show that, although the maximally symmetric vacuum has the same quadratic spectrum, the supersymmetry transformations, the couplings and the scalar potential are parameter dependent. 
We also comment on the possible M-theory uplift and on the meaning of the parameter for the dual gauge theories.
\end{abstract} 


\maketitle

\section{Maximal Supergravity} % (fold)
\label{sec:maximal_supergravity_in_4_dimensions}

The maximally supersymmetric supergravity theory in 4 dimensions has been a fundamental testing ground for our understanding of supersymmetric theories over the years.
For instance, the Scherk--Schwarz supersymmetry breaking mechanism had its first application in the context of the maximal supergravity model obtained by reduction of 11-dimensional supergravity \cite{Scherk:1979zr,Cremmer:1979uq}. 
More recently, new methods for computing loop amplitudes have been developed in the analysis of the quantum perturbative regime of the ungauged theory, building the case for its possible finiteness \cite{Bern:2011qn}.
Finally, using the gauge/gravity correspondence, a SO(8) gauged model \cite{de Wit:1981eq,de Wit:1982ig} with an $N =8$ vacuum has been used to analyze and study possible deformations of the so-called ABJM 3-dimensional conformal theories  \cite{Aharony:2008ug,Aharony:2008gk}.

Actually, $N=8$ supergravity in 4 dimensions comes in two main flavours: the ungauged models, realizing the maximally supersymmetric Poincar\'e algebra, and the gauged theories, where spontaneous supersymmetry breaking can occur.
Here we are mainly concerned with the latter.
Gauged supergravity models are supersymmetric deformations of the ungauged versions by a procedure that couples its vector fields to charges assigned to the other fields according to their transformation properties under the global symmetries of the starting Lagrangian.
From the ungauged model of Cremmer--Julia \cite{Cremmer:1978ds,Cremmer:1979up} one can generate the SO(8) gauged supergravity coming from M-theory compactified on the 7-dimensional sphere \cite{de Wit:1981eq,de Wit:1982ig} or the Scherk--Schwarz gaugings mentioned above \cite{Scherk:1979zr,Cremmer:1979uq}.
However, there are in general very many possible deformations with different gauge groups and also inequivalent models with the same gauge groups.
A first general analysis culminated in the classification of the gaugings embedded in SL(8,${\mathbb R}$) \cite{Cordaro:1998tx}, but it was soon recognized that many more models would escape such classification \cite{Hull:2002cv}.

There are two types of transformations one may use to generate new $N=8$ Lagrangians and hence new massive deformations.
The first one is electric/magnetic duality \cite{Gaillard:1981rj}, mixing the 28 vector fields of the supergravity multiplet and the 28 magnetic duals, which obviously do not appear in the Lagrangian.
Together they transform in the representation \textbf{56} of the U-duality group E$_{7(7)}$ and, although the Lagrangian cannot be invariant under E$_{7(7)}$, the combined equations of motion and Bianchi identities of the vector fields do transform covariantly in the representation \textbf{56}.
Hence (in the ungauged case) the resulting theories are equivalent.
In fact the rigid symmetry group of the Lagrangian is generically only a subgroup of E$_{7(7)}$ and this group is not unique. 
The second set of transformations is related to a larger group, namely Sp(56, ${\mathbb R}$), which determines which gauge fields belonging to the representation \textbf{56} play the role of electric and which ones the role of magnetic gauge fields.
This selects the so-called symplectic frame, which in turn fixes the rigid symmetry group of the ungauged Lagrangian.
Different choices of symplectic frame yield in general different Lagrangians which are not related to each other by local field redefinitions and eventually lead to different gaugings.

% section maximal_supergravity_in_4_dimensions (end)

\section{Embedding Tensor and Duality Orbits} % (fold)
\label{sec:embedding_tensor_and_duality_orbits}

A modern framework to treat all these issues is given by the embedding tensor formalism, introduced in \cite{Cordaro:1998tx,Nicolai:2000sc,Nicolai:2001sv} and developed within the context of the maximal theory in 4 dimensions in \cite{deWit:2002vt,deWit:2005ub,de Wit:2007mt}.
Once the symplectic frame has been chosen and therefore the ordering of the  (electric and magnetic) vector fields $A^M_\mu$, $M=1,\ldots,56$, has been fixed, the embedding tensor $\Theta_{M}{}^\alpha$, gives their coupling to the E$_{7(7)}$ generators $t_{\alpha}$, for instance, via covariant derivatives $D_\mu = \partial_\mu - A_\mu^M \Theta_M{}^\alpha t_{\alpha}$.
From the general analysis of \cite{de Wit:2007mt} we now know that the embedding tensor is fixed by up to \textbf{912} parameters, as this is the only irreducible representation that survives consistency and supersymmetry constraints.
Using such formalism, it is now possible to proceed to a general analysis by which to classify and to construct all massive deformations of $N=8$ supergravity and possibly discuss their vacuum structure \cite{Borghese:2011en,DallAgata:2011aa,Dibitetto:2012ia,Dall'Agata:2012cp}.

In a given symplectic frame, one may expect a one to one correspondence between allowed embedding tensors and gauged supergravity models.
However, most of the theories that share the same gauge group are simply going to be different realizations of the same model, which can be transformed into each other by  U-duality.
For this reason, it is crucial to find an efficient criterion to decide whether two theories are related by such transformations, referring only to the embedding tensor.
For this purpose one can use the techniques that have been developed in the context of supergravity black holes, where the construction of duality invariant quantities depending on the charges allowed to classify inequivalent solutions and to construct the corresponding duality orbits.
Since in the case at hand the black hole charges are replaced by gauging charges, specified by the embedding tensor, we have to find appropriate contractions of the embedding tensor that do not transform under the duality group.
Unfortunately, most of the simple combinations one could think of vanish due to the quadratic constraint \cite{de Wit:2007mt}
\begin{equation}
	\label{quadconstraint}
	\Theta_M{}^\alpha \Theta_N{}^\beta \Omega^{MN} = 0,
\end{equation}
where $\Omega$ is the Sp(56, ${\mathbb R}$) invariant tensor, or their computation is a too demanding task, like for the two-times quartic E$_{7(7)}$ invariant constructed from both the fundamental and adjoint representations.

The way out has been suggested again in the black hole context in \cite{Fre:2011uy}.
In fact, in order to classify duality orbits one may use covariant quantities rather than invariant ones.
Using covariant tensors, one may still extract quantities with constrained transformation properties, differentiating inequivalent expressions.
In particular we will focus on the following quartic tensor classifier \footnote{We could also construct a duality invariant classifier by taking its trace (either using $\eta$ on the upper and lower indices or $\delta$ on the mixed combinations), but unfortunately it is identically vanishing for the case we are interested in.}:
\begin{equation}
	\label{Btensor}
	B_{\alpha \beta}{}^{\gamma \delta} = \Theta_M{}^\gamma \Theta_N{}^\delta \Theta_P{}^\epsilon \Theta_Q{}^\zeta \, d^{MNPQ} \, \eta_{\alpha\epsilon} \eta_{\beta \zeta},
\end{equation}
where $d^{MNPQ}$ is the E$_{7(7)}$ quartic invariant and $\eta$ is the Cartan--Killing metric.
The action of the duality group on (\ref{Btensor}) is non-trivial and well defined by its adjoint indices: $B \to U B U^{-1}$.
This implies that its eigenvalues are not going to change upon the application of a duality transformation and hence they can be used as duality invariant quantities, to classify different embedding tensors.

% section embedding_tensor_and_duality_orbits (end)

\begin{figure*}[t]
	\hspace{-7mm}
 \includegraphics[scale=0.2]{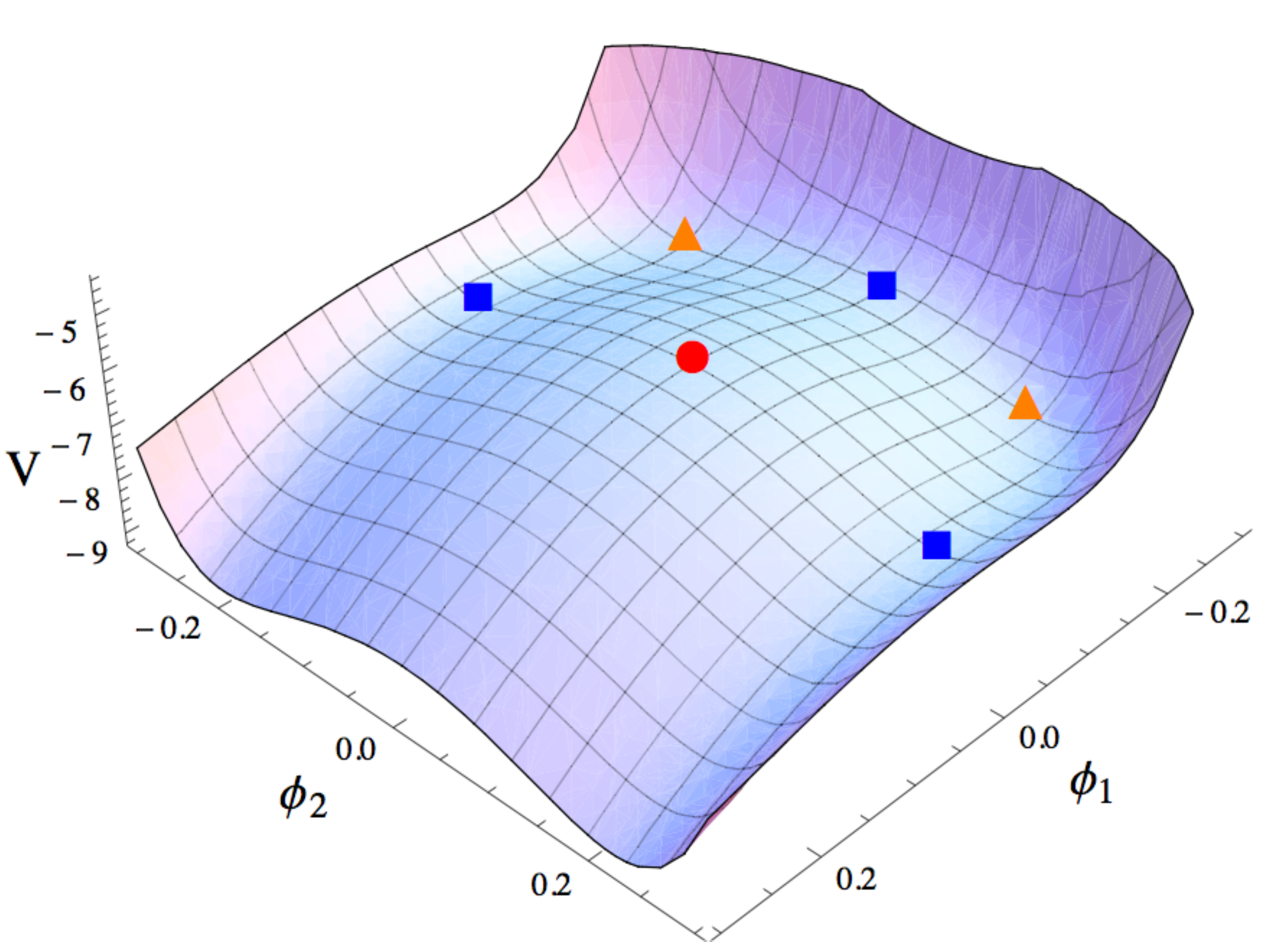} \hspace{27mm}
 \includegraphics[scale=0.2]{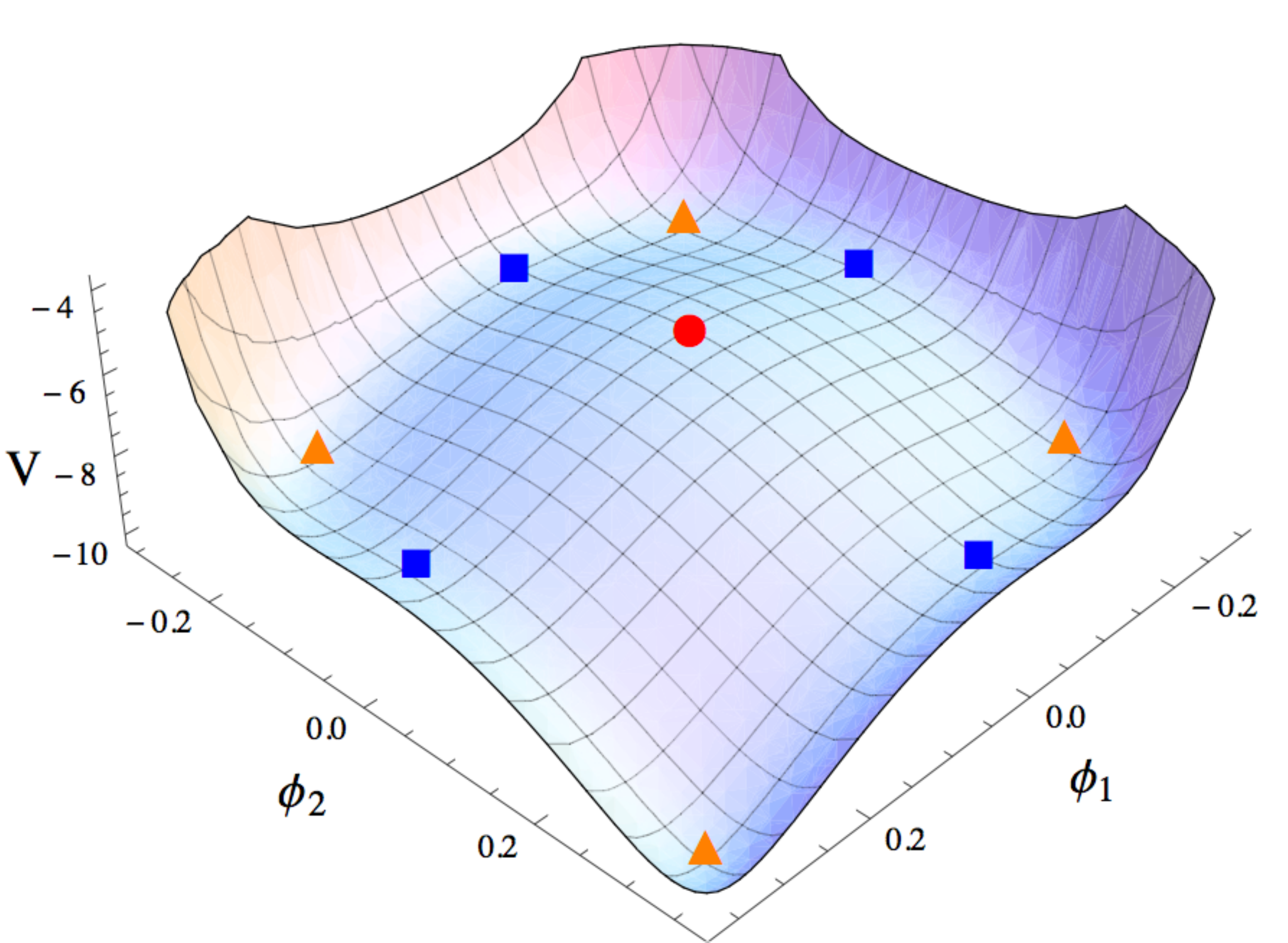}
\caption{\label{fig:vuoti} (Color online) Scalar potential of the G$_2$ truncation for $\omega = 0$ (left) and for $\omega = \pi/8$ (right). The red dot is the SO(8) vacuum, the blue squares are vacua with SO(7) symmetry and orange triangles represent vacua with G$_2$ residual gauge symmetry. New SO(7) and G$_2$ vacua appear with respect to the $\omega = 0$ case.}
\end{figure*}

\section{SO(8)$_c$ Gauging} % (fold)
\label{sec:so_8_gauging}

The gauging of a group $G \subset$ E$_{7(7)}$ requires that the embedding tensor admits at least one singlet in the decomposition of its \textbf{912} under E$_{7(7)} \to G$.
For the SO(8) gauge group one sees that \cite{Andrianopoli:2008ea} 
\begin{equation}
	\mathbf{912} \to 2\times (\mathbf{1}+\mathbf{35}_s +\mathbf{35}_v +\mathbf{35}_c +\mathbf{350}),
\end{equation}
so that actually there are two possible independent tensors specifying its embedding \footnote{One may wonder whether the same argument applies to the 5-dimensional maximal supergravity. One can actually show that this is not the case as the maximal embedding of SU(4) in the SL(2,${\mathbb R}) \times $ SL(6,${\mathbb R}$) decomposition of E$_{6(6)}$ contains only one singlet: $\mathbf{351}\to \mathbf{1}+2 (\mathbf{6}+\mathbf{10}+\overline{\mathbf{10}})+4\cdot \mathbf{15}+\mathbf{20}+\mathbf{45}+\overline{\mathbf{45}}+2\cdot\overline{\mathbf{64}}$.}.
In fact, in the SL(8,${\mathbb R}$) symplectic frame, the electric vector fields transform in the ${\bf 28}$ of $\mathrm{SL}(8,\mathbb{R})$ while the magnetic ones in the ${\bf 28}'$: $A_\mu^M = \{A_\mu^{[AB]}, A_{\mu\,[AB]}\}$, where $A,B=1,\ldots,8$ are indices labelling the fundamental representation of ${\mathfrak sl}$(8,${\mathbb R}$).
The SO(8) gauge group can be obtained by the standard choice \cite{deWit:2002vt}
\begin{equation}
	\label{thetacoupling}
	\Theta_{M}{}^\alpha = \Theta_{AB}{}^C{}_D \propto \delta_{[A}^C \theta^{\vphantom{C}}_{B]D},
\end{equation}
where $\theta_{AB}$ denotes the component of the embedding tensor in the ${\mathbf{36}^\prime}$, which couples the electric vectors to the SL(8,${\mathbb R}$) generators $t_C{}^D$.  
When $\theta$ is positive definite, it defines the SO(8) gauge group and it can be reduced to $\theta_{AB} = \delta_{AB}$ by an SL(8,${\mathbb R}$) transformation.
However, we could also gauge SO(8) by using the magnetic fields and by introducing a second tensor $\xi$ in the $\mathbf{36}$ of SL(8,${\mathbb R}$),  so that \cite{DallAgata:2011aa}
\begin{equation}
	\label{xicoupling}
	\Theta^{AB\,C}{}_D \propto \delta^{[A}_D \xi_{\vphantom{D}}^{B]C},
\end{equation}
and $\xi = c\, \theta^{-1}$ in order to satisfy the quadratic constraint (\ref{quadconstraint}).
This means that, starting from the same ungauged theory, we have a one-parameter family of possible SO(8) gauged supergravity theories, which we will call SO(8)$_c$, depending on the ratio of the couplings (\ref{thetacoupling}) and (\ref{xicoupling}).
Most of these models are going to be dual to each other and we can explicitly check their equivalence by using the tensor classifier (\ref{Btensor}).

In order to simplify computations and avoid redundancies due to the remaining SL(8,${\mathbb R}$) invariance, we take the embedding tensor in the form: 
\begin{equation}
	\Theta_{AB}{}^C{}_D = \delta_{[A}^C \delta^{\vphantom{A}}_{B]D}, \quad 	\Theta^{AB\,C}{}_D = c\, \delta^{[A}_D \delta_{\vphantom{A}}^{B]C},
\end{equation}
with all the other components vanishing.
We can then compute the eigenvalues of the $B$ classifier, which are
\begin{equation}
	{\rm eig} \, B = \left\{8806 \times \lambda_1, 35 \times \lambda_2, 35 \times \lambda_3, 35 \times \lambda_4 \right\},
\end{equation}
where
\begin{eqnarray}
&&	\lambda_1 = 0, \quad \lambda_2 = 18\, c^2,\\
&& \lambda_3 = \frac92\, (c^2-1)^2, \quad \lambda_4 = -\frac92\, (c^2+1)^2.
\end{eqnarray}
One could still perform and overall rescaling of $\Theta$, which amounts to a simple redefinition of the coupling constant, and tune the values of such eigenvalues to obtain equivalent gaugings, but we can see that this rescaling would not affect the ratios $\lambda_i/\lambda_4$.
We can also see that the spectrum of these ratios is invariant under the maps $c \to -c$, $c \to 1/c$ and $c \to \frac{c-1}{c+1}$, so that we can argue that we have inequivalent gaugings for $c \in [0, \sqrt2 -1]$.
In fact, at this point it is more efficient to parameterize the 1-parameter family of inequivalent gaugings by
\begin{equation}
	\label{embomega}
	\Theta_{AB}{}^C{}_D = \cos \omega\, \delta_{[A}^C \delta^{\vphantom{A}}_{B]D}, \ \  	\Theta^{AB\,C}{}_D = \sin \omega\, \delta^{[A}_D \delta_{\vphantom{A}}^{B]C},
\end{equation}
so that $\lambda_2/\lambda_4 = - \sin^2 (2 \omega)$ and $\lambda_3/\lambda_4 = - \cos^2 (2 \omega)$.
Hence, for $\omega = 0$ we recover the original gauging, for $\omega = \pi/2$ we obtain the dual one, constructed using the magnetic vectors, and inequivalent ones come in the range $\omega \in [0,\pi/8]$.
We remark that having different eigenvalues of the tensor $B$ is a \emph{sufficient} condition to claim that the corresponding theories are inequivalent, because duality transformations cannot change their value.

\emph{These results imply that the SO(8) gauged supergravity built in \cite{de Wit:1981eq,de Wit:1982ig}, also obtained as the compactification of M-theory on $S^7$, is not unique.}

% section so_8_gauging

\section{Couplings of SO$(8)_c$} % (fold)
\label{sec:couplings_of_so_8__c_}

All the inequivalent models discussed above have an $N=8$ vacuum with a negative cosmological constant (which is the same in the conventions of (\ref{embomega})) and obviously the quadratic spectra around such vacua coincide.
However, higher order couplings change, as expected for inequivalent models.
We will now show explicitly some of the SO(8)$_c$ couplings and compute their dependence on the parameter, with a special emphasis on the scalar potential, which now shows a different spectrum of vacua according to the parameter's choice.

For the sake of clarity we restrict the analysis of the potential to the   G$_2$-invariant sector of the scalar fields.
It is known that for $c=0$ one finds one $N=8$ vacuum with SO(8) symmetry, two parity conjugated vacua with $N=0$ and SO(7)$^-$ residual symmetry, another $N=0$ vacuum with SO(7)$^+$ symmetry, self-conjugated under parity, and two parity conjugated $N=1$ vacua with G$_2$ symmetry \cite{Warner:1983vz}.
The G$_2$-invariant truncation contains two scalar fields $\vec \phi = (\phi_1,\phi_2)$ and  the potential can be written as the sum of three pieces
\begin{equation}
	\label{G2potential}
	V(\vec \phi) = A(\vec \phi) - \cos (2 \omega) f(\phi_1,\phi_2) - \sin (2 \omega) f(\phi_2,\phi_1),
\end{equation}
where (in the following $x \equiv e^{|\vec \phi|}$)
\begin{eqnarray}
	A(\vec \phi) &=&\textstyle \frac{(1+ x^4)^3}{64 |\vec \phi|^4 \,x^{14}}\left[4(1+ x^4)^2(1-5 x^4+ x^8)(\phi_1^4+\phi_2^4) \right. \nonumber\\[1mm]
	&+& \left.\phi_1^2 \phi_2^2 (1+ 4x^4-106 x^8+4x^{12}+x^{16})\right],
\end{eqnarray}
which is an even function of $\phi_1$ and $\phi_2$ and symmetric in their exchange, and 
\begin{equation}
	\begin{array}{rl}	
	f(\phi_1,\phi_2) &= \frac{(-1+ x^4)^5 \,\phi_1^3}{64 |\vec \phi|^7 \,x^{14}}\left[4(1+5 x^4 + x^8)\phi_1^4 + \right.\\[2mm]
	&+ \left. 7 (1+ 6 x^4 + x^8) \phi_1^2 \phi_2^2 + 7 (1+x^4)^2 \phi_2^4\right],
	\end{array}
\end{equation}
which is odd in the first argument and even in the second.
Three symmetry operations leave the scalar potential invariant: 
\begin{equation}
	\label{symmetriespot}
	\left\{\begin{array}{l}
	\omega \leftrightarrow - \omega \\[2mm]
	\phi_2 \leftrightarrow - \phi_2
 	\end{array}\right.\,, \ \ 
	\left\{\begin{array}{l}
	\omega \leftrightarrow \omega + \frac{\pi}{2} \\[2mm]
	\vec\phi \leftrightarrow - \vec\phi
 	\end{array}\right.\,, \ \ 
	\left\{\begin{array}{l}
	\omega \leftrightarrow \omega-\frac{\pi}{4} \\[2mm]
	\phi_1 \to \phi_2 \\[2mm]
	\phi_2 \to -\phi_1
 	\end{array}\right..
\end{equation}
The first one results from a parity-related symmetry, while the last two result from $\mathrm{E}_{7(7)}$-duality transformations.
Altogether this implies that we get inequivalent potentials only in the expected range $\omega \in [0,\pi/8]$.
In fact, depending on the parameter $\omega$ the scalar potential exhibits a different number of vacua, as shown in Fig.~\ref{fig:vuoti}.
The $\omega = 0$ case corresponds to the usual truncation of the scalar potential that keeps the SO(8) vacuum (although seemingly unstable, all the masses satisfy the Breitenlohner--Freedman bound), the SO(7)$^{\pm}$ vacua and the G$_2$ ones.
When $\omega \neq 0$ a new SO(7) and new G$_2$ vacua appear.
In fact, not only the number of vacua changes when $\omega \neq 0$, but also the value of their cosmological constant, as can be seen by looking at figure~\ref{fig:vuoti}.
In particular, we computed the ratio of the value of the cosmological constant of the various vacua in the two potentials with respect to that of the $N=8$ vacuum in the center.
The result is an $\omega$-dependent function, different for each one of the vacua.

A crucial ingredient in any gauged supergravity theory is given by the expression of the shifts of the supersymmetry transformation rules.
Among other couplings, they determine the fermion masses as well as the scalar potential.
In particular, the gravitino shift, determining also the gravitino masses, is diagonal in the G$_2$ truncation discussed above:
\begin{eqnarray}
A_1 &=& {\rm diag}\{\mu_1, \mu_1, \mu_1, \mu_1, \mu_1, \mu_1, \mu_1, \mu_2\},
\end{eqnarray}
where
\begin{eqnarray}\textstyle
\mu_1 &=& -\frac{1}{2|\vec\phi|^4} 
\left[ e^{-i\omega}(\phi_1-i\phi_2)^2\,{\rm ch}^3|\vec\phi|\,h_1^-(\phi_1,\phi_2)\right.\nonumber\\
&+&\left. e^{i\omega}\,|\vec\phi|\,(\phi_1+i\phi_2)\,{\rm sh}^3|\vec\phi|\,h_1^+(\phi_1,-\phi_2)\right],\\[.5em]
\mu_2 &=& -\frac{1}{2|\vec\phi|^4}
\left[e^{-i\omega}\,(\phi_1+i\phi_2)^2\,{\rm ch}^3|\vec\phi|\,h_2^-(\phi_1,\phi_2)\right.\nonumber\\
&+&\left.e^{i\omega}\,|\vec\phi|^{-3}(\phi_1+i\phi_2)^5\,{\rm sh}^3|\vec\phi|\,h_2^+(\phi_1,-\phi_2)\right]
\end{eqnarray}
and
\begin{equation}
\begin{array}{rcl}
h_1^{\pm}(\phi_1,\phi_2) &=& {\rm ch}(2|\vec\phi|)\,\left( 6\phi_1^2+8i\,\phi_1\phi_2-6\phi_2^2 \right)\\[.5em]
&\pm&(3+{\rm ch}(4|\vec\phi|))\,\left(2\phi_1^2+3i\,\phi_1\phi_2-2\phi_2^2\right),
\end{array}
\end{equation}
\begin{equation}
	\label{finaf}
\begin{array}{rcl}
h_2^{\pm}(\phi_1,\phi_2) &=& h_1^{\pm}(\phi_1,\phi_2)+8i\,\phi_1\phi_2\,{\rm ch}(2|\vec\phi|).
\end{array}
\end{equation}

In the  $\mathrm{SL}(8,\mathbb{R})$-frame the gauge group defined by
the chosen embedding tensor does not have an electric action on the
vector field strengths and their duals.
However, as proved in \cite{deWit:2005ub}, we can always choose an electric frame for any gauging.
Starting from (\ref{embomega}), we get to the electric frame by using the symplectic rotation:
\begin{equation}
{\cal E} = \left(\begin{array}{cc}
\cos\omega & \sin\omega \\
-\sin\omega & \cos\omega
\end{array}\right) \otimes 1_{28}.
\end{equation}
When we perform this transformation, the kinetic terms of the vector fields and, more in general, the non-minimal couplings of the scalars to the vector fields get a non-trivial dependence on the parameter $\omega$.
${\cal E}$ has no effect on the SO(8) generators but brings back the gauge connection to be electric. 
Hence in the ${\cal E}$-symplectic frame $\Theta_M{}^\alpha$ and $X_{MN}{}^P$ are identical to the ones of \cite{de Wit:1981eq,de Wit:1982ig}.
Still, the couplings in (\ref{G2potential})-(\ref{finaf}) retain their $\omega$ dependence, because the embedding of E$_{7(7)}$ in Sp(56,${\mathbb R}$) changed and hence the explicit form of the E$_{7(7)}$ generators that are not in SO(8) also changed.

% section couplings_of_so_8__c_ (end)

\section{Comments} % (fold)
\label{sec:comments}

The techniques we described can be applied in full generality to any family of gaugings of maximal supergravity sharing the same gauge group.
In particular most of our discussion carries over to the SO$(p,q)$ gaugings discussed in \cite{Hull:1984wa,Cordaro:1998tx,DallAgata:2011aa}.

Having a 1-parameter family of inequivalent SO(8) gauged supergravity theories poses some interesting puzzles from both the string theory point of view as well as for the gauge/gravity correspondence.
It is known that the compactification of M-theory on $S^7$ can be consistently truncated to the usual 4-dimensional gauged SO(8) theory of \cite{de Wit:1981eq,de Wit:1982ig,de Wit:1986iy,Nicolai:2011cy}.
Is there a reduction procedure leading to the SO(8)$_c$ theories?
It is clear that changing frame is related to a different choice of the fundamental vector fields remaining in the dimensional reduction process.
For instance, while the electric gauge fields come from the reduction of the metric, the SO(8)$_c$ theories involve magnetic gauge fields whose higher-dimensional origin is not obvious.
Also, the maximal supersymmetric vacuum of the SO(8) model provides the gravity dual to the ABJM theory \cite{Aharony:2008ug}, but what could the extra parameter correspond to?
Given that also the SO(8)$_c$ theories have an $N=8$ vacuum with the same spectrum as that of the SO(8) model, one should expect that the deformed theories correspond to dual field theories with the same chiral ring, same 2-point functions, but different higher-point functions.
Actually, \cite{Aharony:2008gk} proposes a generalization of the ABJM theories related to M2-branes probing ${\mathbb C}^4/{\mathbb Z}_k$, whose near-horizon limit is $AdS_4 \times S^7/{\mathbb Z}_k$.
The orbifold action is such that for $k=2$ the short supergravity spectrum remains untouched and therefore the 4-dimensional truncations would look the same at the quadratic level.
Furthermore, having an orbifold allows for the introduction of discrete torsion, leading to two different theories, according to the amount of discrete torsion introduced \footnote{We thank J.~Maldacena and D.~Jafferis for suggesting this possibility.}.
Obviously the discrete torsion is a discrete parameter and cannot capture all the SO(8)$_c$ deformations, but this may well be an artifact of the classical 4-dimensional supergravity models, where we did not consider any quantization condition.
The authors of \cite{Aharony:2008gk} propose three different models that may be related to our supergravity theories: a model with gauge group U($N$)$_1 \times $U$(N)_{-1}$, one with gauge group U$(N)_2 \times $U$(N)_{-2}$ and another one with gauge group U$(N+1)_{2} \times $U$(N)_{-2} $ (and the parity-reversed U$(N+1)_{-2} \times $U$(N)_{2} $).
The ${\mathbb Z}_2$ orbifold projection does not affect the massless spectrum and the first two models are also parity invariant.
The third one is not parity invariant, but is invariant under the combination of parity and the duality transformation U$(N+l)_k \times $U$(N)_{-k} \to $U$(N+|k|-l)_{-k} \times $U$(N)_k$, where $k$ is related to the ${\mathbb Z}_k$ orbifold projection and $l$ is the number of fractional branes.
According to our previous discussion, we therefore suggest the identification
\begin{equation}
	\omega = \frac{\pi}{4} \frac{l}{k}, \qquad k =1,2, \quad l \leq k.
\end{equation}
By this relation between the parameters the first two models are identified, while the third one is invariant under the exchange of scalar fields in the representations $\mathbf{35}_v$ and $\mathbf{35}_s$.
This exchange is actually the result of the combined application of parity and of the duality transformation (4.19) in \cite{DallAgata:2011aa}.
We also point out that $\omega$ survives the truncation to $N =6$ and therefore it would be interesting to analyze that case, too.

These are surely very interesting problems, which we plan to explore in the future.

% section comments (end)

\vskip 0.1in

{\it Acknowledgments --}
We would like to thank M.~Bianchi, F.~Bigazzi, B.~de Wit, S.~Ferrara, T.~Fischbacher, D.~Jafferis, J.~Maldacena, F.~Marchesano, A.~Marrani, A.~Zaffaroni and F.~Zwirner for helpful discussions. This work is supported in part by the ERC Advanced Grants no. 226455, ``SUPERFIELDS'', by the European Programme UNILHC (contract PITN-GA-2009-237920), by the Padova University Project CPDA105015/10, by the MIUR-PRIN contract 2009-KHZKRX and by the MIUR-FIRB grant RBFR10QS5J.

\end{document}